\documentclass[useAMS]{mn2e}
\usepackage{rotate}
\usepackage{graphicx}

\def\be{\begin{equation}}
\def\ee{\end{equation}}

\input{tcilatex}

\begin{document}

\title[Double features and curvature radiation in pulsars]{Do double
features in averaged pulsar profiles decipher the nature of their radio emission?%
}
\author[J. Gil and G. Melikidze]{Janusz A. Gil$^1$\thanks{%
E-mail: jag@astro.ia.uz.zgora.pl} and George I. Melikidze$^{1,2 }$\thanks{%
E-mail: gogi@astro.ia.uz.zgora.pl} \\
%EndAName
$^1$J. Kepler Institute of Astronomy, University of Zielona G\'ora , Lubuska
2, 65-265 Zielona G\'ora, Poland\\
$^2$ Abastumani Astrophysical Observatory, Ilia State University, 2A Kazbegi ave., Tbilisi GE-0160,
Georgia}
\date{Accepted . Received ; in original form }
\maketitle

\begin{abstract}
An interesting paper has recently been published claiming that the long-sought Rosetta Stone needed
to decipher the nature of pulsar radio emission has been finally identified as the double features
in averaged pulsar profiles. The authors argue that highly symmetric bifurcated features are
produced by a split-fan beams of extraordinary-mode curvature radiation emitted by thin microscopic
streams of magnetospheric plasma conducted by a very narrow bundle of magnetic field lines. We
examined arguments leading to these intriguing conclusions and found a number of flaws. At least
one of them is fatal, namely there is not enough available energy within such thin microscopic
plasma streams. Using an elementary pulsar physics we show that if the stream is so thin that its
emission can reveal the signatures of elementary radiation mechanism, then the energy deficit tends
to be severe and reaches a few to several orders of magnitude (depending on the actual efficiency
of converting the available kinetic energy of relativistic charged particles into the coherent
radio emission). We are certain that the answer to the question contained in the title of this
paper is definitely negative.
\end{abstract}

\pagerange{\pageref{firstpage}--\pageref{lastpage}} \pubyear{2010}

\label{firstpage}


\section{Introduction}

Almost half a century passed since the discovery of pulsars, but yet no agreement has been reached
concerning the actual mechanism for their radio emission (of generation of their observed radio
emission). The exceedingly high brightness temperature that can be deduced from the observed flux
densities undoubtedly implies that the pulsar radiation must be emitted coherently. Generally, the
pulsar radio emission can be generated by means of either a maser-like or the coherent curvature
mechanism (Ginzburg \& Zheleznyakov 1975; Kazbegi, Machabeli \& Melikidze 1991; Melikidze, Gil \&
Pataraya 2000, Paper I hereafter). Without any doubt this radiation is emitted in a strongly
magnetized electron-positron plasma well inside the light cylinder. Once the waves are generated in
the emission region, in the propagation region they naturally split into the ordinary and
extraordinary waves corresponding to the normal modes of the strongly magnetized plasma (see e.g.
Arons \& Barnard 1986; Lominadze et al. 1986). The ordinary waves are polarized in
the plane of the wave vector ${\bmath k}$ and the local magnetic field ${%
\bmath B}$ and their electric field has a component along both ${\bmath k}$ and ${\bmath B}$.
Therefore, they interact strongly with plasma particles (causing charge-separation along field
lines), and thus encounter strong difficulty in escaping from the magnetosphere. On the other hand,
the extraordinary waves are linearly polarized perpendicularly to the plane containing both
${\bmath k}$ and ${\bmath B}$ vectors and thus they cannot separate charges along ${\bmath B}$. As
a result the extraordinary mode can propagate freely through the magnetospheric plasma and escape
to the interstellar medium. Many observational constrains on the emission altitude unambiguously
suggest that the emitted radiation detaches from the magnetosheric plasma at altitudes $r$ being
less than 10\% of the light cylinder radius $R_{LC}=2\pi /Pc$ (e.g. Blaskiewicz, Cordes \&
Wasserman 1991; Kijak \& Gil 1997). It is worth emphasizing that this also holds for the
millisecond pulsars (see Figure 3 in Kijak \& Gil 1998)

Therefore, from a theoretical point of view the bulk of the observed pulsar radiation originates
when the extraordinary plasma waves escape from the magnetosphere. There exists also strong
observational indication that the extraordinary mode is dominant in pulsar radiation. Indeed, Lai,
Chernoff \& Cordes (2001) found that the Vela pulsar emits radio waves polarized predominantly in
the direction perpendicular to the plane of dipolar magnetic field lines. In fact, they were able
to demonstrate convincingly that in the fiducial phase\footnote{ The fiducial phase corresponds to
the fiducial plane, which contains both the rotation and the magnetic axes as well as the
line-of-sight (that is the wave-vector ${\bmath k}$)} the radiation is polarized perpendicularly to
the plane of the dipolar magnetic field lines. This argument was extended by Gil, Lyubarsky \&
Melikidze (2004, Paper II hereafter ) to every phase within the pulse window, using the fact that
the mean position angle swing in this pulsar follows the rotating vector model (RVM hereafter;
Radhakrishnan \& Cooke,1969; Johnston, van Straten, Kramer \& Bailes, 2001). This means that the
radiation observed at a given longitude is polarized perpendicularly to the plane of dipolar
magnetic field lines, along which the sources of this emission are moving. It is worth noting that
the above conclusion concerns the average polarization. The analogous problem related to the
instant emission observed in single pulses was recently discussed by Mitra, Gil \& Melikidze (2009,
Paper III hereafter).

The open question is which of the two possible mechanisms of coherent radio emission: maser--like
or curvature radiation, is responsible for generation of the observed pulsar radiation? Let us keep
in mind that it must be the extraordinary mode (to escape freely from the magnetosphere) polarized
perpendicularly to the planes of dipolar magnetic field lines (to satisfy observational
polarization constrains). The most natural candidate is the coherent curvature radiation, as it is
the only mechanism that distinguishes planes of magnetic field lines (source trajectories).
Recently Mitra, Gil \& Melikidze (2009) found strong arguments to help distinguish between the
alternative mechanisms. They analyzed highly polarized (nearly 100\%) single pulses in a number of
strong pulsars and argued that they can be produced only by the extraordinary mode excited in the
magnetospheric plasma by means of the coherent curvature radiation. In fact, they found that
position angle variations in subpulses precisely follow the RVM--like mean position angle swing.
This is exactly what is expected to be produced by curvature radiation in a plasma, whose
polarization is perpendicular to the magnetic field line planes. The maser-like emission generates
fast swings of instant position angle in subpulses, incompatible with the RVM (see Paper III for
details).

It would certainly be desirable to find additional and independent observational evidences to
support the coherent curvature radiation being the mechanism for generation of pulsar radio
emission. Recently, Dyks, Rudak \& Rankin (2007; DRR07 hereafter) and Dyks, Rudak \& Demorest
(2010; DRD10 hereafter) claimed that the double symmetrical features (called the bifurcated
components; BFC hereafter) and notches (absorption features) observed in averaged profiles of some
pulsars (e.g. their Figs. 1 and 2) carry the crucial information able to decipher the nature of the
observed radio emission. The idea was to fit these features with the elementary emission pattern of
selected radiation mechanisms: the parallel acceleration beam and the curvature radiation beam.
Although we do not believe that the actual pulsar radiation mechanism can be identified from the
analysis of the properties of average waveforms alone, we do not intent to argue with the approach
itself. Rather, we intend to verify the arguments and claims of DRD10 using different and
independent methods.

Early attempts were unsuccessful and it turned out that the fit of the parallel acceleration beam
model proposed by DRR07 "was the wrong idea". Most recently DRD10 ultimately gave up on the
parallel acceleration mechanism and concentrated on the curvature radiation. They claimed that the
observed pulsar radio emission was the coherent curvature radiation (more precisely the component
polarized perpendicularly to the planes of magnetic field lines). At first we were glad to see
published conclusions strongly supporting our results obtained by means of completely independent
arguments and methods (Papers I, II and III). Later we found a number of flaws in considerations of
DRD10. We decided to present and discuss these flaws in this paper, as we were afraid that
misleading and incorrect arguments of DRD10 can rather harm the idea of the coherent curvature
radiation as the mechanism of pulsar emission than promote it. Some readers may get an impression
that the long standing problem of the physical mechanism of the pulsar emission has just been
solved. Unfortunately this is not true and still more work is to be done in this field.

First of all, the considerations of DRD10 are based on the single particle curvature radiation
mechanism in vacuum, while it is well known that pulsar radiation must be emitted coherently in the
magnetized electron-positron plasma. Apparently, DRD10 assumed implicitly that a hypothetical
coherence mechanism (which they did not specify) would just reconstruct the single particle vacuum
case. However this is not true. As we demonstrate in this paper the single particle vacuum model
cannot even be used as a first approximation of coherent curvature radiation in a magnetized
plasma.

\section{Double features and notches}

As already mentioned, DRD10 ignored realities of the excitation of radio waves and their
propagation in the pulsar magnetospheric plasma. All their arguments were based on the formalism of
the single particle curvature radiation excited and propagated in vacuum. They strongly concluded
that the observed pulsar radiation was the coherent curvature radiation, although without any
justification for the coherency part. They apparently believed that their arguments could be
directly applied to the realistic pulsar environment. For example they state: "The bunching-induced
coherency (e.g. due to the two-stream instability, Ruderman \& Sutherland 1975; RS75 hereafter)
seems to have less problems with the quasi-isotropic amplification of the non-coherent beam." This
statement apparently mixes two independent problems, namely: creation of bunches and generation of
coherent radiation. The bunches formed naturally by linear electrostatic waves (which DRD10 seem to
be referring to) cannot provide any emission, because the characteristic time-scale of such
bunching is too short compared with the time-scale of curvature radiation (see Paper I for
details). On the other hand, the RS75 mechanism corresponds to vacuum, where there would be no
problem with generation of electromagnetic waves with both polarizations (provided they could be
emitted). Thus, the problem DRD10 seems to be facing is how to dump the parallel mode of curvature
radiation, which in vacuum is 7 times stronger than the perpendicular one (e.g. Jackson 1975). This
happens naturally in magnetized plasma but at the same time properties of the escaping curvature
radiation change their characteristics with respect to the vacuum case (Paper II). Thus, the latter
cannot be used as a credible diagnostic tool to unravel the pulsar radiation mechanism.

To the best of our knowledge the only efficient and physically self--consistent mechanism of bunch
creation in the pulsar magnetospheric plasma is the spark associated soliton model developed in
Paper I. This model uses the two-stream instability\footnote{It is worth emphasizing that this is
the only plasma instability that can occur in the near pulsar magnetosphere (e.g. Asseo \&
Melikidze (1998).} exclusively to generate the longitudinal (non-electromagnetic) plasma waves. The
actual bunching is caused by the nonlinear evolution of these plasma wave-packets and formation of
charged relativistic solitons capable of emitting coherent curvature radiation. The influence of
the ambient magnetized plasma on both generation and propagation of this radiation was studied in
Paper II, where it was shown that only the extraordinary mode (with polarization perpendicular to
the planes of magnetic field lines) can escape from the pulsar magnetosphere. Following Paper II we
will use the point-like approximation model of soliton bunches, making our arguments quite general,
i.e. independent of the actual bunching mechanism. We will demonstrate that characteristic
properties of the coherent curvature radiation in the pulsar magnetospheric plasma are quite
different from that of the textbook vacuum case explored by DRD10. We will examine an influence of
the plasma on the frequency dependence of the bifurcation angle of the BFC. Additionally, we will
check if there is enough kinematic energy (the uppermost limit to the radiation energy) to power
the BFC within the thin plasma stream model considered by DRD10.

\subsection{Fitting the bifurcated component}

The most important feature considered by DRD10 is a bifurcated component (BFC) in the mean profile
of PSR J1012+5307 (see their Fig.2). This feature has a high (although not perfect) degree of a
mirror symmetry and DRD10 argue that it reflects a pure morphology of an extraordinary mode of
curvature radiation \footnote{ Strictly speaking they mean the component of curvature radiation
polarized perpendicularly to the plane of source motion (i.e. in the plane of curved magnetic field
line), as neither ordinary nor extraordinary mode exists in vacuum. Both parallel and perpendicular
polarization components of curvature radiation in vacuum have the feature of the extraordinary
mode, which is defined by the absence of the electric field component along the wave-vector.}.
Indeed, waves polarized perpendicularly to the plane of the charge motion are not emitted along the
instant velocity vector (see lower panel of Figure 3 in Paper II, where it is clearly seen that
there is no curvature radiation emitted at and near the local magnetic field direction). DRD10 in
their Section 3.2.2 adopt a strong assumption concerning the origin of BFC feature, namely that
"the emitter has a form of thin and elongated plasma stream that emits the curvature radiation
mainly in the extraordinary (orthogonal) mode". However, their Eqs.(3) and (4) corresponding to the
vacuum case contain two electromagnetic components, both of which should reach the observer.
Moreover, in contrast to their assumption of missing the parallel component, its power is about 7
times higher than the power of the perpendicular mode. In order to dump the parallel component one
needs a plasma environment, where the corresponding equations for the power of polarized radiation
are significantly different from those in vacuum (see Paper II). DRD10 used the vacuum formalism to
fit the curvature radiation to the BFC of PSR J1012+5307, while the proper approach would be to fit
the coherent curvature radiation in plasma using a general formalism developed in Paper II (see
next Section for details). Moreover, DRD10 fit both the parallel acceleration beam and curvature
radiation beam models to the BFC profile and conclude that the latter is slightly better, although
the former cannot be excluded. However, they found that only the curvature radiation model can
match the data simultaneously at two frequencies. In the next section we examine this problem under
the proper treatment including the ambient plasma influence.

\subsection{Frequency dependence of the bifurcation angle}

\begin{figure}
\includegraphics[scale=0.8]{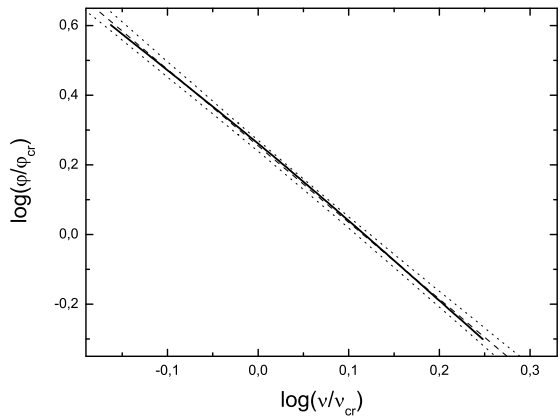}
\caption{Frequency dependence of the opening angle of the extraordinary mode
of coherent curvature radiation in the pulsar magnetospheric plasma (thick
line). The formal power-low fit is represented by the dashed line and the $%
95\%$ upper and lower limits are represented by dashed lines.}
\label{figBFC}
\end{figure}

As mentioned above, DRD10 argue that the important property of BFC supporting the model of
curvature radiation is the frequency dependence of the bifurcation angle (angular separation
between component peaks; see their Figs. 8 and 9) $\Delta _{\mathrm{bfc}}\propto \nu
_{\mathrm{obs}}^{-0.35}$. Interestingly, the value of this exponent is close to $1/3=0.333$, which
indeed follows from the properties of the single particle curvature
radiation in vacuum (provided that $\Delta _{\mathrm{bfc}}=1/\gamma $ and $%
\nu _{\mathrm{obs}}\propto \gamma ^{3}$, e.g. Jackson 1975). This would be an impressive feature
supporting the curvature radiation model if one could be certain that the ambient plasma does not
change the value of the bifurcation exponent. However, as we will demonstrate below, this is not
the case. Moreover, even DRD10 admit that "the exponent of $1/3$ is not ubiquitous among pulsars".
In fact, as one can see in their Fig. 6 PSR J1012+5307 is the only pulsar having the bifurcation
exponent close to $1/3$ (it is also worth noting that the measured value $0.35$ is not equal to
$1/3$ even within the error bars marked in the figure). In realistic theory of the coherent
curvature radiation emitted and propagating in a pulsar plasma this exponent can differ from $1/3$
and/or $0.35$. As an example we explored the formalism developed in Paper II, which corresponds to
the general case of curvature radiation emitted by the point-like (smaller than the emitted
wave-length) charged bunch/soliton moving relativistically along curved magnetic field lines. Only
the extraordinary mode polarized perpendicularly to the planes of field lines can reach the
observer. We calculated an opening angle between the local magnetic field and the direction at
which the maximum power of this mode is emitted in a plasma as a function of frequency (for
illustration see the upper panel of Figure 3 in Paper II, where it is clearly seen that the
extra-ordinary mode is missing towards the direction of the local magnetic field). The results are
presented in Fig.\ref {figBFC}, where on the horizontal axis is the normalized observational
frequency $\nu /\nu _{cr}$ (where $\nu _{\mathrm{cr}}=7.2\times 10^{9}\gamma ^{3}/{%
\ \rho }$ [Hz], and $\gamma $ is the Lorentz factor of the source of curvature radiation moving
relativistically along the trajectory with the radius of curvature $\rho $), while on the vertical
axis is the normalized angle $\varphi /\varphi _{\mathrm{cr}}$ between the direction of maximum
power emission and the local magnetic field vector (where $\varphi _{\mathrm{cr} }=1/\gamma $).

Unlike the vacuum case, the bifurcation frequency dependence is not exactly a power-law like for
the curvature radiation in a plasma (thick solid line in Fig. \ref{figBFC}). However, when the
formal power-law fit was applied we obtained $\varphi /\varphi _{\mathrm{cr}}\propto \left( \nu
/\nu _{\mathrm{cr}}\right) ^{-a}$, where $a=0.45\pm 0.01$. This value of bifurcation exponent is
far from $1/3$ that can be derived for the vacuum case. Moreover, it does not depend on the actual
bunching mechanism (e.g. soliton model), that is any point-like emitter of curvature radiation will
give the BFC exponent equal to $0.45$ in the pulsar plasma. Also, the elementary pattern of the
curvature emission in a plasma is different from that of emitted in vacuum. Moreover, the spectrum
of the soliton curvature radiation differs from that of single particle even if both are calculated
in vacuum (see Fig. 4 in Paper I), not to mention the influence of the plasma environment (Paper
II). The quality of fit of the curvature radiation pattern to the BFC profile in a plasma will
certainly be much worse than in the vacuum case. Thus, in our opinion there is no reason to believe
that neither the formal fit itself nor the exponent value ($0.35$) in the frequency dependence of
BFC found by DRD10 in PSR J1012+5307 is the signature of the actual radiation mechanism, the
curvature radiation in particular. If indeed there was a possibility to detect and resolve the
elementary feature of the curvature radiation (from a single source or a number of sources flowing
along a narrow bundle of field lines and emitting in the same plane) then the bifurcation exponent
should be about $0.45$ and not about $0.35$. Moreover, as demonstrated in the next section there is
not enough power within such a narrow bundle to provide the observed BFC luminosity, which is even
more fatal for the model proposed by DRD10. It is worth emphasizing here that we do believe
coherent curvature radiation to be the actual pulsar radiation mechanism, but for completely
different reasons than those presented by DRD10 (see Papers I, II and III). We think that the BFC
feature in PSR J1012+5307 represents normal components in the complex mean profile of this pulsar
and the measured bifircation/separation index is close to $1/3$ by accident. There are many factors
influencing a value of this index besides the intrinsic radiation mechanism, such as:
radius-to-frequency mapping, radius of curvature varying across the emission region, size of the
emission region, organization of the emission region (conal versus patchy) and geometry of pulsar
emission.  (e.g. Gil \& Krawczyk 1996, Gil et al. 2002).

\subsection{Energy considerations\label{EnC}}

As already mentioned DRD10 used the textbook formalism of single particle curvature radiation in
vacuum and implicitly assumed that all properties of the observed coherent radiation will be
exactly the same when considered properly in the pulsar magnetospheric plasma (although they never
expressed this explicitly). We conjecture that this assumption concerned the energy problem as
well, which however DRD10 did not consider at all. We can make simple estimates of \textit{the most
upper limits} of a possible emission power, and compare them with the observed radio luminosities.
The mean flux density from PSR J1012+5307 at 1.4 GHz is about 3 mJy (ATNF database), which for the
distance $d=0.52$ kpc translates into the radio luminosity $L_{r}\sim 6\times 10^{27}$ erg s$^{-1}$
(Eq.(3.14) in Lorimer \& Kramer, 2005). Judging from the mean profiles shown in Figure 2 of DRD10
it is reasonable to assume that the BFC feature contains not less than about $10\%$ of the total
energy associated with the whole pulse profile. The same must hold for the emitted power, so we can
assume that $L_{\mathrm{BFC}}\sim 0.1L_{r}=6\times 10^{26}$ erg s$^{-1}$. The highest available
energy source is determined by the spin down power $L_{\mathrm{SD}}\sim 4\times 10^{31}{%
\dot{P}_{-15}}/P^{3}$ erg s$^{-1}$, which for PSR J1012+5307 with $P=0.0053$ s and
$\dot{P}_{-15}=1.7\times \times 10^{-5}$ gives $L_{\mathrm{SD}}\sim 4.7\times 10^{33}$ erg
s$^{-1}$. We can now calculate the so-called pulsar kinematic luminosity $L_{\mathrm{kin}}$, which
is the power carried by charged particles accelerated within the pulsar inner gap (see Section 6.2
in Paper II). This luminosity can be expressed as $L_{\mathrm{kin}}=\gamma _{
\mathrm{pr}}m_{e}c^{3}n_{\mathrm{GJ}}S_{\mathrm{PC}}$ erg s$^{-1}$, where
$\gamma_{\mathrm{pr}}<5\times 10^{6}$ is the "primary" Lorentz factor of
electrons (or positrons) leaving the acceleration region ("polar gap"), $n_{%
\mathrm{GJ}}=1.4\times 10^{11}({\dot{P}}_{-15}/P)^{0.5}$ cm$^{-3}$ is the
Goldreich-Julian (1969) number density and $S_{\mathrm{PC}}=6.6\times
10^{8}P^{-1}$ cm$^{2}$ is the canonical polar cap surface area. For the
parameters of this pulsar $n_{\mathrm{GJ}}=8\times 10^{9}$ cm$^{-3}$ and $S_{%
\mathrm{PC}}=1.3\times 10^{11}$ cm$^{2}$ and thus $L_{\mathrm{kin}}<10^{32}$
erg s$^{-1}$, which is few percent of $L_{\mathrm{\ SD}}$ (as should be
expected in general). The radiation efficiency $\eta $ of the observed radio
emission referred to the kinematic pulsar luminosity is $\eta =L_{r}$/$L_{%
\mathrm{kin}}$=$6\times 10^{27}/10^{32}=6\times 10^{-5}$, which is quite typical for radio pulsars.
Below we argue that such low efficiency should be also expected in conversion of the particle's
kinematic luminosity into the power of coherent curvature radiation. Indeed, we can estimate the
kinetic energy flux of the coherently emitting bunches as $\tilde{L}_{\text{kin}}=\gamma
m_{e}c^{3}n_{\text{GJ}}S$. If we assume that the entire $\tilde{L}_{\text{kin}}$ is converted into
the radio emission (i.e. $L_{r}=\tilde{L}_{\text{kin}}$), then the efficiency would be
$\eta_{\text{CR}} =\tilde{L}_{\text{kin}}/L_{\text{kin}}=\gamma /\gamma _{\text{pr}}$. For the
typical values of Lorentz factors $\gamma =400$ and $\gamma _{pr}=10^{6}$ of the secondary and the
primary plasma, respectively (see Papers I and II for details), we get $\eta_{\text{CR}} =4\times
10^{-4}$ as a maximum possible efficiency estimation. More realistically, if only a part of
$\tilde{L}_{\text{kin}}$ is converted into $L_{r}$, then the efficiency is much lower, close to
typical value measured in radio pulsars, i.e. $\eta_{\text{CR}} \sim \eta=6\times 10^{-5}$.

For a convenience of further considerations we will now introduce the
surface density of kinematic luminosity, which can be defined as $L_{0}=L_{
\mathrm{kin}}/S_{\mathrm{PC}}$. For the parameters of this pulsar $%
L_{0}<10^{21}$ erg s$^{-1}$ cm$^{-2}$. According to DRD10 the BFC feature is emitted via curvature
radiation of sources flowing within a very narrow flux tube of magnetic field lines (see Figure 12
in their paper), although its actual cross--section is not specified. On one hand, this flux tube
should be broad enough to carry much more than $L_{\mathrm{BFC} }=6\times 10^{26}$ erg/s in the
particle flux.
\begin{figure}
\includegraphics[scale=0.5]{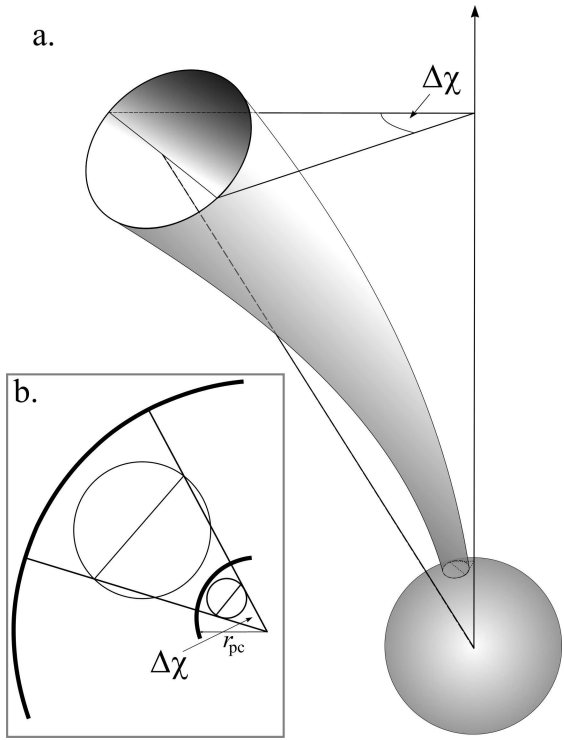}
\caption{a. The sketch of the flux tube associated with the bifurcated component (BFC). b. The top view of the polar
cap and the base of the BFC flux tube with a divergence $\Delta\chi$.}
\label{figFlux}
\end{figure}
On the other hand, to reveal signatures of elementary emitters it should be narrow enough so that
its divergence $\Delta \chi =\varepsilon /\gamma $ is much smaller than the opening angle of
curvature radiation $1/\gamma $. Thus, the auxiliary dimensionless  parameter $\varepsilon <<1$.
Now we can roughly estimate the required cross-section of the flux tube satisfying the above
condition. We will use a projection along the dipolar field lines from the radiation region onto
the stellar surface (see Fig. 2 for a schematic sketch). Near the edge of the polar cap we can
write for the dimension of the base of the BFC flux tube and for its
cross--section $\Delta = \Delta\chi r_{\mathrm{pc}}$ and $S=\pi r_{\mathrm{pc%
}}^2\Delta\chi^2=S_{\mathrm{pc}}\Delta\chi^2$, respectively. Here $r_{\mathrm{%
pc}}$ and $S_{\mathrm{pc}}$ are the radius and surface area of the canonical polar cap,
respectively. Now we can write that
\begin{equation}
S=S_{\mathrm{PC}}\frac{\varepsilon ^{2}}{\gamma ^{2}}=1.3\times 10^{11}\frac{%
\varepsilon ^{2}}{\gamma },  \label{Cross}
\end{equation}%
and the associated kinematic luminosity carried along the BFC flux tube is $%
L=L_{0}S=10^{32}\varepsilon ^{2}\gamma ^{-2}$ erg/s. In order to estimate the value of $\gamma $
let us note that curvature radiation has a maximum emissivity at the frequency $\nu _{m}=2\times
10^{9}\gamma ^{3}/R$ Hz, where $R>10^{7}r^{0.5}$ cm is the radius of curvature of the magnetic
field lines and $r$ is the emission altitude (see e.g. RS75). This frequency has to be about 1 GHz
and thus, assuming reasonably that $r>3\times 10^{6}$ cm, we obtain $\gamma >225$. Therefore the
kinematic luminosity of BFC flux tube $L<2.4\times 10^{27}\varepsilon ^{2}$ erg/s. We still need to
estimate the value of $\varepsilon $ which should be much less than unity. Let us note that even if
we adopt as the most upper limit $\varepsilon \simeq 0.3$, we get the kinematic luminosity $L<
L_{BFC}=3.6\times 10^{26}$ erg s$^{-1} $.
It is worth remembering that the extraordinary (perpendicular polarization) mode carries only $%
1/7 $-th part of the total emitted power, so the maximum efficiency of this mode is about 15\%.
This leads to the most upper limit $L\simeq 5\times 10^{25}$ erg/s, much less than $L_{BFC}=6\times
10^{26}$ erg s$^{-1}$. In practice the actual efficiency is much smaller, but even assuming this
absolutely unrealistic maximum efficiency there is not enough luminosity to power the BFC feature.

The above estimate corresponds to unrealistic parameter values: $\gamma _{%
\mathrm{pr}}=5\times 10^{6}$, $\gamma =220$ and $\varepsilon =0.3$. For more
realistic values $\gamma _{\mathrm{pr}}=10^{6}$, $\gamma =400$ and $%
\varepsilon =0.1$ (e.g. Paper I) one obtains $L<10^{24}$ erg s$^{-1}$ $<<L_{BFC}$. Taking into
account that this kinematic power still does not include the efficiency of curvature emission we
can see that the power deficit reaches several order of magnitudes. Once this efficiency is taken
into account then one obtains
\begin{equation}
L_{r}^{\mathrm{BFC}}<\frac{\gamma }{\gamma _{\mathrm{pr}}}L=2.5\times
10^{25} \frac{\varepsilon ^{2}}{\gamma }.
\end{equation}
The left--hand side of this equation represents the radio luminosity of the
BFC feature, which should account for about $L_{BFC}=6\times 10^{26}$ erg s$%
^{-1}$. For the first and the second set of values of $\varepsilon ,\gamma $ and $\gamma
_{\mathrm{pr}}$ used above one obtains $L_{r}^{\mathrm{BFC}}$ equal to $0.03$ and $6.2$ times
$10^{20}$ erg s$^{-1}$, respectively, which is much less than $L_{BFC}$ in both cases. Moreover,
the inequality sign in the above equation is related to simplifying assumptions that we used,
namely that charge density inside the coherently emitting bunch is equal to the local
Goldreich-Julian (1969) value and the entire kinetic energy is converted into the radio emission,
while in fact this is an upper limit.

In summary, the model of the BFC emission presented by DRD10 has an energy deficit amounting to
several orders of magnitude and it cannot be balanced by any means. This conclusion is independent
of the actual pulsar radiation mechanism, i.e. it holds for any bunching mechanism (including the
soliton model) in the pulsar plasma, the parallel acceleration beam, maser-like emission, etc. The
BFC feature can be emitted from the bundle of field lines base of which covers at least 10\% of the
polar cap area, once again irrespective of the actual radiation mechanism.

We believe that BFC feature in PSR J1012+5307 is just a normal part of the multi-component average
profile of this pulsar. What seems to distinguish it from the rest of the profile is its high
degree of symmetry. However, such kind of symmetry is nothing extraordinary among the pulsar mean
profiles. Moreover, one cannot exclude that the overwhelming symmetry of the BFC feature in PSR
J1012+5307 (see Figs. 3 and 4 in DRD10) will disappear at different frequencies, which often
happens in double component profile pulsars (some tendency to such change is visible in Figs. 8 and
9 of DRD10). Perhaps the best example of extremely symmetric profile is the case of PSR J0631+1036
(see fig.2 in Weltevrede et al. 2008). Both the overall waveform and separation between the inner
components in this profile is about the same as in the BFC of PSR J1012+5307. However, the
excellent polarimetry available for the former pulsar indicates that its profile is emitted close
to the fiducial plane$^1$ and the separation/birfucation of several degrees of longitude is
impossible for Lorentz factors $\gamma $ being of the order of 100 (necessary for the
characteristic frequency of curvature radiation to be in the radio band). In PSR J1012+5307 DRD10
solved this problem by postulating "the sightline cuts through plasma streams", but in PSR
J0631+1036 the plasma streams are apparently cut more centrally. Otherwise, the double/birfurcated
components look alike in both pulsars.

\section{Conclusions}
In the last paragraph of their paper DRD10 state "We conclude that the long-sought Rosetta Stone
needed to decipher the nature of pulsar radio emission has finally been identified as double
features in averaged pulse profiles". This statement, based on incomplete set of evidence, is
simply not right. The BFC feature that DRD10 proclaimed the "Rosetta Stone" cannot be emitted in
the way to play a role of the latter. There is not enough available energy source to power the
features that could reveal signatures of the elementary pulsar emission, no matter whether in
single pulses or in average profiles.

The huge energy deficit is the most serious problem of DRD10 model. It appears that the kinematic
power begins to be balanced if the base of the bundle of magnetic field lines carrying the plasma
stream associated with the BFC feature covers at least 10 \% of the polar cap surface area,
irrespective of the actual radiation mechanism. Of course, such stream is too wide to reveal the
physical properties of the elementary emission, as there must be many sources of the coherent
pulsar radiation distributed over its cross-section.

Another major problem of DRD10 is the frequency dependence of bifurcation of the two components of
BFC feature (see their Figs. 8 and 9). DRD10 claim that this feature is produced by a split-fan
beam of extraordinary-mode curvature radiation emitted along the sufficiently thin plasma stream.
With no emissivity of the extraordinary-mode in the plane of field lines carrying this stream, the
observed radiation should be bifurcated, with a bifurcation angle (angular distance between the two
apparent components) being dependent on frequency. Using a very high quality observational data
DRD10 found that $\Delta _{\mathrm{bfc}}\propto \nu _{\mathrm{obs}}^{-a}$, where $a=0.35\pm 0.01$.
They claimed that this was fully consistent with a low-frequency curvature radiation for which
$a=1/3$. First of all, the observed value is not equal to $1/3$ even within error bars (see their
Fig. 6), which is of course not a big problem. More serious problem is related to the fact that
DRD10 used a textbook expression for the single particle curvature radiation emitted and propagated
in vacuum. Although DRD10 never say it explicitly they implicitly suggest that the single particle
vacuum curvature radiation is a very good approximation of the actual pulsar radiation mechanism
(or in other words the influence of plasma on the generation and propagation of pulsar radiation
can be neglected). While the single-particle radiation is not a bad model for a coherent radiation
by a small point-like bunches, a vacuum approximation is absolutely not acceptable. We found that
for the curvature radiation emitted and propagated in pulsar plasma (only the extraordinary-mode
can leave the magnetosphere) the value of bifurcation exponent $a=0.45\pm 0.01$. This should be the
measured value of the bifurcation exponent if indeed it was possible to detect and resolve the
elementary feature of the curvature radiation in the pulsar radio emission. Interestingly, there is
one point in Fig.6 of DRD10 with $a=0.42\pm 0.025$ (J0437), but we think that this is by pure
accident. Indeed, a "normal" separation exponents for components in the complex pulsar profiles can
have any value between about $0.25$ and $0.7$, depending on different geometrical conditions (e.g.
Gil \& Krawczyk 1996, Gil et al. 2002 and Table 6 in DRD10). We believe that the BFC feature is not
different from the rest of the profile of PSR J1012+5307. Apparently, this is an almost aligned
rotator and the polar cap is "seen" for almost an entire pulsar period. It would be extremely
interesting to obtain a single pulse data, but this will probably be possible only with a future
generation radiotelescopes. One important prediction is the following: if indeed the BFC feature in
the profile of J1012+5307 represents the signatures of extraordinary mode of the curvature
radiation then all single pulses beneath this component should also be bifurcated and look alike
average emission. If it turns out not to be the case, then BFC feature is nothing extraordinary,
just a normal macroscopic double component in multicomponent profile of this pulsar.

The two major problems discussed above are fatal for the DRD10 model, especially the missing energy
problem. Besides them there are few minor problems and we would like to mention two of them here.
The natural consequence of DRD10 model is an elongated fan structure of pulsar beams. Such a
structure is difficult to reconcile with the observed rates of occurrence of interpulses (IP),
including double pole (DP) and single pole (SP) cases. Observational data show that in the entire
pulsar sample there are about 2 and 1 percent of the former and the latter, respectively. Every
statistical study of pulse profile widths that includes interpulses indicates that pulsar beams
must be nearly circular (Gil \& Han 1996, Kolonko, Gil \& Maciesiak 2004, Weltevrede \& Johnston
2008, Keith et al. 2010). Indeed, any significant deformation of beam circularity spoils the
expected IP statistics. In particular, the elongated fan beams produce too many interpulses as
compared with observations (e.g. Gil \& Han 1996).

Another minor problem is the following. The intrinsic beam-width of the BFC feature is less than
one degree, while the observational bifurcation is about 8 degrees of longitude. To achieve the
apparent broadening by a factor of 10 or so DRD10 must use extreme values of geometrical
observational angles (small cut angles) and/or substellar radii of curvature of magnetic field
lines. This doesn't seem very likely, although such geometrical situation cannot be excluded.

We did not touched the phenomenon of notches at all, which seem to be bifurcated absorption
features occurring in a few pulsars. According to DRD10 the physics of notches is identical to that
of BFC and the only difference is geometrical in nature. Namely, the double absorption features are
produced by beam of the extraordinary mode of curvature radiation, when it is eclipsed by the thin
plasma stream. We understand that this beam is analogous to the one associated with the BFC feature
discussed by DRD10 and thus the proposed origin of notches is subject to the same criticism as that
of BFC related emission. However, the BFC is not visible and the only role the thin plasma stream
plays is to eclipse a normal pulsar radio emission. This can happen due to geometrical (the thin
beam emission misses the line-of-sight) or energetic (the beam is not producing the coherent radio
emission) reasons. With this latter interpretation the origin of notches proposed by DRD10 can be
perhaps considered as a viable model. However, it cannot be done without problems. One problem is
how to pin firmly a thin plasma stream to the polar cap surface. The other, more serious problem is
that the extraordinary mode of curvature radiation is not supposed to be absorbed by any kind of
magnetospheric plasma.

The criticism of DRD10 model presented in this paper does not change our opinion that the pulsar
observed radiation is really the extraordinary mode of the coherent curvature radiation emitted and
propagated in the magnetospheric plasma (Papers I, II and III). One is tempted to say that DRD10
came to the right conclusion for the wrong reasons.

\section*{Acknowledgments}

This paper was partially supported by research Grants N N 203 2738 33 and N N 203 3919 34 of the
Polish Ministry of Science and Education. GM was partially supported by the GNSF grant ST08/4-442.
We thank Boe Lewandowski for critical reading of the manuscript.

\end{document}